\documentclass[runningheads]{llncs}
\usepackage[T1]{fontenc}
\usepackage{graphicx}
\usepackage{booktabs}
\usepackage{multirow}
\usepackage{amsmath}
\usepackage{amssymb}
\usepackage{algorithmic}
\usepackage[ruled,linesnumbered]{algorithm2e} 
\usepackage[misc,geometry]{ifsym}
\newcommand{\etal}{et~al.}

\usepackage{color}
\begin{document}
\title{The Loop Game: Quality Assessment and  Optimization for  Low-Light Image Enhancement}
\titlerunning{Quality Assessment and Optimization for Low-Light Image Enhancement}
%

\author{Danni Huang\inst{1}\textsuperscript{$\ddagger$} \and
Lingyu Zhu\inst{2}\textsuperscript{$\ddagger$} \and 
Zihao Lin\inst{1} \and 
Hanwei Zhu\inst{3} \and 
Shiqi Wang\inst{2}\textsuperscript{(\Letter)} \and 
Baoliang Chen\inst{1}\textsuperscript{(\Letter)} }
\authorrunning{D. Huang and L. Zhu}
%
\institute{Department of Computer Science, South China Normal University 
\email{blchen6-c@my.cityu.edu.hk}\\
\and
Department of Computer Science, City University of Hong Kong
\email{shiqwang@cityu.edu.hk}\\
\and
College of Computing and Data Science, Nanyang Technological University\\
}

\maketitle             

\renewcommand{\thefootnote}{\fnsymbol{footnote}} 
\footnotetext[1]{Danni Huang and Lingyu Zhu contributed equally. Baoliang Chen and Shiqi Wang are the corresponding authors.} 

\begin{abstract}
There is an increasing consensus that the design and optimization of low light image enhancement methods need to be fully driven by perceptual quality. With numerous approaches proposed to enhance low-light images, much less work has been dedicated to quality assessment and quality optimization of low-light enhancement. In this paper, to close the gap between enhancement and assessment, we propose a loop enhancement framework that produces a clear picture of how the enhancement of low-light images could be optimized towards better visual quality. In particular, we create a large-scale database for QUality assessment Of The Enhanced LOw-Light Image (QUOTE-LOL), which serves as the foundation in studying  and developing objective quality assessment measures. The objective quality assessment measure plays a critical bridging role between visual quality and enhancement and is further incorporated in the optimization in learning the enhancement model towards perceptual optimally. Finally, we iteratively perform the enhancement and optimization tasks, enhancing the low-light images continuously. The superiority of the proposed scheme is validated based on various low-light scenes.

\end{abstract}
\section{Introduction}
Low light image enhancement fills in the gap between the prevalence of image acquisition devices and poor quality photos under low-light conditions, and provides the meaningful surrogate for expensive hardware solutions. Typically,  
the enhancement algorithms aim to improve the visual quality from multiple perspectives due to a series of visual quality degradations by inadequate and unbalanced lighting, such as low visibility, undesirable color cast, and intensive noise.
Pioneering efforts are dedicated to the reconstruction from the low-light one to the normal illumination, including the Histogram Equalization (HE) based approaches \cite{arici2009histogram} and 
Retinex theory-based approaches \cite{fu2016weighted,jobson1997multiscale}. However, only enlightening the image without inducing any other image priors could inevitably lead to amplified noise. Recently, the data-driven approaches based upon deep-learning have achieved substantial breakthroughs. In particular, they rely on the large-scale database with paired low/high-quality images, such that the enhancement mappings can be learned in a fully-supervised manner.

\begin{figure}[t]
\begin{minipage}[b]{1.0\linewidth}
  \centering
  \centerline{\includegraphics[width=1\linewidth]{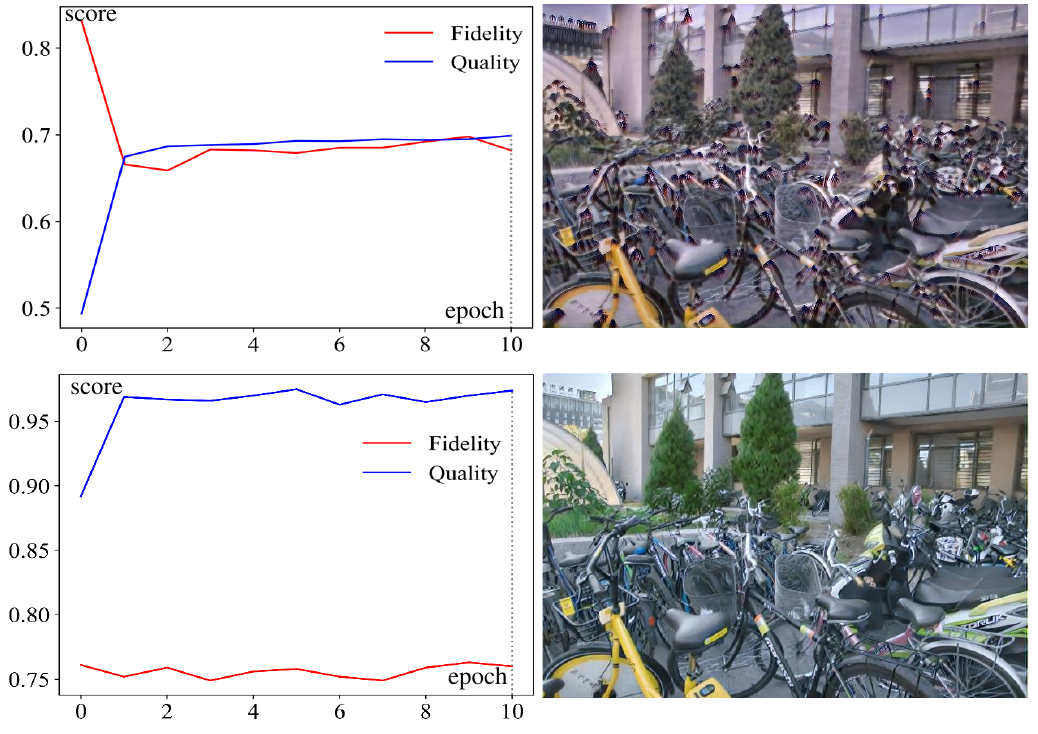}}
\end{minipage}
\caption{Optimization results with different fidelity and quality measures. First row: optimized by SSIM \cite{wang2004image} (fidelity) loss and NIMA \cite{talebi2018nima} (quality) loss. Second row: optimized by our proposed fidelity and quality measures in the first iteration. We divide the output scores of NIMA  by 10 for unification.}
\label{fig:opti}
\end{figure}

One of the major challenges of fully-supervised based methods lies in the optimization process that is rarely guided by perceptual quality models. 
Despite the fact that many fidelity measures have been committed to optimizing low-light image enhancement methods~\cite{zhu2022enlightening,wei2018deep,yang2020fidelity}, there are still underlying limits. When it comes to \textit{error visibility approaches}, mean absolute error (MAE) and mean squared error (MSE) are poorly correlated with the human vision system. Though the \textit{structural similarity index} was motivated by human perception~\cite{wang2004image}, the gray image assumption made it unsuitable for optimization in the color domain. \textit{Learning-based methods} have recently played an important role in perceptual image optimization, but those methods are prone to overfitting the training data, and the optimized images may contain extra artifacts~\cite{ding2021comparison}. Furthermore, the quality of the reference image limits the supervised low-light image enhancement methods. As a result, to develop a method that can generate visually pleasing results beyond the ground truth, some works attempted to learn the low-light enhancement model using unsupervised learning strategy~\cite{jiang2021enlightengan}. Though such methods have great potential in exploiting the scene statistics of natural images, it is still quite questionable that the learned statistics could well reflect human perception without clear guidance. Talebi~\etal provided another solution that combined the fidelity loss with a no-reference quality assessment (NR-IQA) model dubbed NIMA~\cite{talebi2018nima}, resulting in a considerable quality improvement. The fidelity loss (SSIM~\cite{wang2004image}) was employed to approximate the ground truth, while the quality loss (NIMA~\cite{talebi2018nima}) was used to monitor image style. However, when we applied the loss to the low-light image enhancement task, the addition of NIMA~\cite{talebi2018nima} has not achieved a perfect result.
There are several reasons behind this. First, 
the NIMA quality model lacks enough generalization capability due to the large domain gap between its training set and the enhanced low-light image. Second, the fidelity loss and quality loss may conflict with each other when they are simultaneously optimized, which is verified by their scores changes as shown in Fig~\ref{fig:opti}. As such, the quality measure may not always convey the right information to the enhancement quality,  resulting in unexpected distortions (\textit{e.g.}, biased color and residual noise).
Although numerous NR-IQA models have been proposed ~\cite{zhu20242afc,chen2025debiased,chen2024deep,9452150,9894272,9447183,10687355}, they are typically not designed for low-light image enhancement, which may result in an incomplete capture of the true distribution of distortions in low-light images.


\begin{table*}[]
\centering
\caption{Performance comparisons of classical IQA methods on the tailored TID2013 \cite{ponomarenko2015image} database.}
\small
\begin{tabular}{c|ccccc|cc}
\toprule
\multirow{2}*{Methods} & NIQE & BRISQUE & \multirow{2}*{PSNR}  & SSIM  & VIF   & MSSSIM  \\
     & \cite{mittal2012making} &\cite{mittal2012no}&&\cite{wang2004image}&\cite{sheikh2006image}&\cite{wang2003multiscale}\\
\midrule
PLCC & 0.2437 & 0.2195 & 0.6313 & 0.6127 & 0.8052 & 0.8049 \\
SRCC & 0.2600 & 0.1566 & 0.6578 & 0.5919 & 0.7084 & 0.7833 \\

\midrule
    \multirow{2}*{Methods} & FSIM  & GMSD~  & MAD   & IWSSIM & AWDS  & \multirow{2}*{Fused} \\
    &\cite{zhang2011fsim}&\cite{xue2013gradient}&\cite{johnson2016perceptual}&\cite{wang2015patch}&\cite{ding2021comparison}\\
\midrule
PLCC & 0.8521 & 0.8103 & 0.7908 & 0.7996 & 0.8332 & \textbf{0.8643} \\
SRCC & 0.8013 & 0.7835 & 0.7165 & 0.7119 & 0.7538 & \textbf{0.8182} \\
\bottomrule
\end{tabular}
\label{tab:tidres}
\end{table*}

In this paper, we address those issues via closing the gap between quality assessment and enhancement in a loop manner. 
In particular, we first contribute a new dedicated database consists of diverse images and different enhancement models, and assign each image with a pseudo-Mean Opinion Score (pseudo-MOS) by the score-mapping of several reliable FR-IQA models. The database construction aims to develop a low-light specific NR-IQA measure. To solve the dilemma between fidelity measure and quality measure, we disentangle an image into the fidelity and quality by the signal decomposition and construct a fidelity compatible quality measure based on the NR-IQA model.
Finally, the loop between quality assessment and enhancement is preformed, through which, the generalization capability of the NR-IQA model can be further improved by continuously updated samples. 
Extensive experimental results demonstrate this loop-closing framework successfully achieves a large leap in terms of visual quality and guarantees better quality of the resulting images after each loop. 
Our main contributions are as follows,
\begin{itemize}
    \item 
    We create a large-scale database termed as QUality assessment Of The Enhanced LOw-Light Image (QUOTE-LOL), consisting of 290 image contents and 10 enhancement methods. In total,  2,900 images as well as  their pseudo-MOSs are created, providing a reliable way for measuring and optimizing the visual quality of the enhanced images. 
    
    \item We develop an NR-IQA model based on the statistics of deep features, aiming for   the image style modulation. 
    The proposed quality measure is then incorporated with our fidelity measure, which could improve the visual quality and preserve image content simultaneously. The high generalization capability is further verified on an unsupervised enhancement framework. 
    \item 
    We propose the first loop solution for low-light image enhancement. The proposed joint loss guides the enhancement network to achieve high-quality results, possessing a potential even superior to the reference, and the enhanced images, in turn, facilitate the quality measure. The game between quality evaluation and quality optimization is performed iteratively, building the  closed loop for continuous quality improvement.
\end{itemize}

\begin{figure*}[h]
\includegraphics[width=\textwidth]{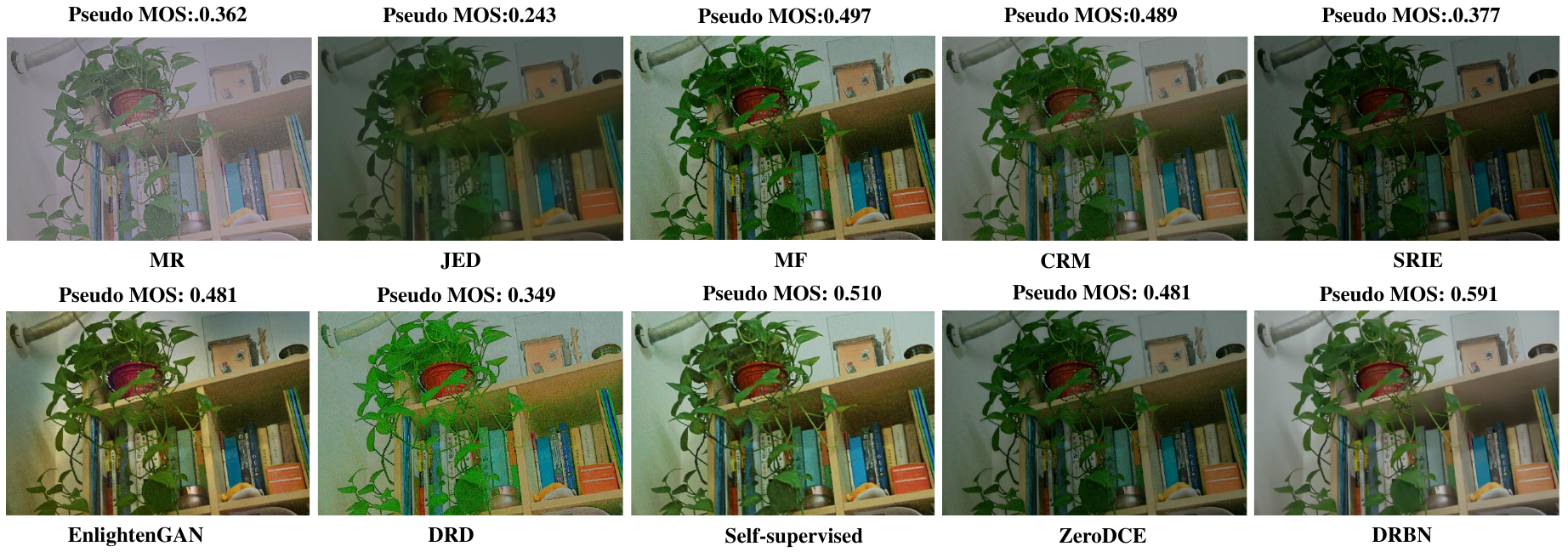}
\caption{Sampled enhanced low-light images with their corresponding pseudo-MOSs in QUOTE-LOL database.}
 
\label{fig:Samples}

\end{figure*}

\section{Closing the Loop of Quality Assessment and Optimization  for Low-Light  Enhancement }

\subsection{Motivation}
Our goal is to enhance the low-light image into a normal-light one with both signal fidelity and perceptual quality.
Previous works usually learn an enhancement network supervised by the reference (normal-light)  images.
However, as indicated in \cite{duanmu2021quantifying}, the fidelity measures between enhanced image and its reference do not always reflect the perceptual quality change.
Namely, beyond the fidelity measurement, an NR-IQA model should also be introduced for final quality optimization.

As we discussed before, a reliable and valid NR-IQA model should: 1) be generalized to reflect the quality changes and provide useful feedback continuously during the enhancement.
2) be compatible with the fidelity loss, namely the profiting of quality should not be paid at a cost of fidelity. 
For the first issue, we create a low-light specific database (QUOTE-LOL) for NR-IQA model learning, aiming to capture and model the intrinsic perceptual characteristics especially for low-light images. 
Besides, by feeding our IQA model with new samples at each loop, its generalization capability can be further improved, leading to continuous quality improvement.
For the second issue, inspired by the SSIM and its related works~\cite{wang2015patch,wang2004image}, we decompose an image patch $\mathbf{x}$ into five components:
\begin{equation}
\begin{aligned}
\mathbf{x} &=\left\|\mathbf{x}-\mu_{\mathbf{x}}\right\|_2 \cdot \frac{\mathbf{x}-\mu_{\mathbf{x}}}{\left\|\mathbf{x}-\mu_{\mathbf{x}}\right\|_2}+\mu_{\mathbf{x}} \\
&=\left\|\tilde{\mathbf{x}}\right\|_2 \cdot \frac{\tilde{\mathbf{x}}}{\left\|\tilde{\mathbf{x}}\right\|_2}+\mu_{\mathbf{x}} \\
&=c_{\mathbf{x}}\cdot \mathbf{s}_{\mathbf{x}}+\mu_{\mathbf{h}}+\mu_{\mathbf{s}}+\mu_{\mathbf{v}},
\end{aligned}
\end{equation}
where $\mu_{\mathbf{x}}$ denotes the mean
value of the patch and we further decompose it into HSV color space \textit{i.e.,} hue ${\mu}_{\mathbf{h}}$, saturation ${\mu}_{\mathbf{s}}$ and lightness ${\mu}_{\mathbf{v}}$. The $c_{\mathbf{x}}$ and $\mathbf{s}_{\mathbf{x}}$ represent the contrast and structure components
of $\mathbf{x}$, respectively. By such decomposition, the 
${\mu}_{\mathbf{h}}$ and $\mathbf{s}_{\mathbf{x}}$ can be regarded as the content (fidelity) information of an image while  the $\mu_{\mathbf{s}}$, $\mu_{\mathbf{v}}$ and  $c_{\mathbf{x}}$ control the image style (quality).

Following this vein, we propose a fidelity and quality compatible joint loss for low-light image optimization. In particular, the structure and hue information is maintained by our fidelity measure and  the style is modulated by our NR-IQA model. In the following subsections, we will introduce our database construction first, then the loop between quality assessment and quality optimization is presented. 
\begin{figure*}[ht]
\begin{center}
\includegraphics[width=1.0\textwidth]{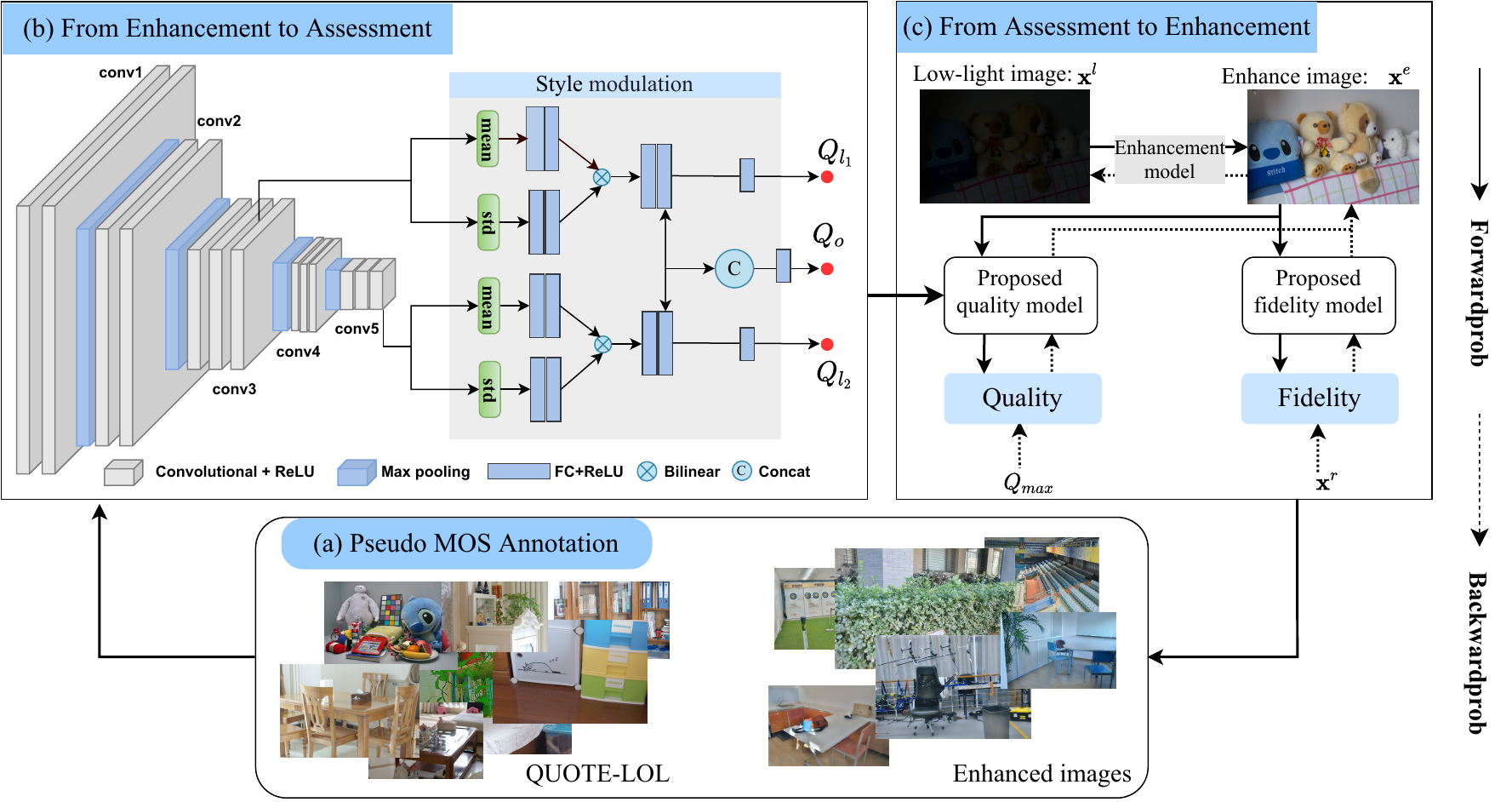}
\caption{
Summary of the proposed scheme which closes-up the loop between quality assessment and enhancement for low-light images.
a) \textit{Pseudo-MOSs annotation} for QUOTE-LOL database and the latest enhanced images. 
b) A deep-learning based IQA measure dedicated to the enhanced images, bridging the gap \textit{from enhancement to assessment};
c) The optimization of enhancement models with the guidance of the IQA model is explored to fill in the gap \textit{from assessment to enhancement}.
{The knowledge transfer from the updatable database to the IQA model and from the IQA model to the enhancement model close the gap between enhancement and assessment.}
}
\label{fig:arc}
\end{center}
\end{figure*}
\subsection{Database Construction}
We create a dedicated database QUOTE-LOL which contains 2,900 enhanced low-light images and corresponding quality labels. Those images are generated by ten  methods including both traditional and deep learning based schemes with various image contents. Based on the developed large-scale database, we demonstrate that a well-behaved IQA algorithm and readily-deployed optimization pipeline can be further developed to ultimately improve the performance of the image enhancement. In particular, we adopt the images in the publicly accessible LOw-Light database (LOL) \cite{wei2018deep} as the source for database construction. 

The LOL database is acquired from realistic and diverse scenes, consisting of 689 low/normal-light training pairs and 100 low/normal-light testing pairs. In the LOL, 247 and 43 low-light images are selected by the contents from the training and testing sets in LOL, respectively. 
Subsequently, we apply ten classical low-light image enhancement methods, including Camera Response Model (CRM) \cite{ying2017new}, Joint Enhancement and Denoising Method (JED) \cite{ren2018joint}, Multiple Fusion (MF) \cite{fu2016fusion},  Multiscale Retinex (MR) \cite{jobson1997multiscale}, Simultaneous Reflectance and Illumination Estimation (SRIE) \cite{fu2016weighted}, EnlightenGAN \cite{jiang2021enlightengan}, Deep Retinex Decomposition (DRD) \cite{wei2018deep},  ZeroDCE \cite{guo2020zero}, Self-supervised \cite{zhang2020self} and Deep Recursive Band Network (DRBN) \cite{yang2020fidelity} for enhancement. In particular, EnlightenGAN and ZeroDCE are the pioneering unsupervised methods for low-light image enhancement.

As shown in Fig.~\ref{fig:Samples} and the supplementary materials, these methods inevitably introduce a wide variety of quality degradation in the enhancement process, including blurring, noise contamination, contrast reduction, detail loss, over-exposure, under-exposure and color shift. More importantly, the noise could even be amplified. These distinct artifacts deliberately or unintentionally introduced with different distortion patterns  are prone to quality issues, and present a stark contrast compared with the synthetic distortions.

To study the quality of the enhanced images and behaviors of low-light enhancement models, inspired by the database construction in~\cite{wu2020end}, we explore 11 classical IQA models, from which three reliable IQA models are selected  for a pseudo-MOS generation. In particular, we first tailor the TID2013 database and collect a subset that only contains the possible distortion types introduced by enhancement, including various types of noise, blur, intensity shift, contrast  as well as change of saturation. The performance comparison of 11 classical IQA models is shown in Table~\ref{tab:tidres}. Without loss of generalization, we then adopt a four parameters  linear mapping function to combine the three selected FR-IQA models (FSIM, IWSSIM, AWDS), 
\begin{equation}\label{map}
\mathrm{Q}=\lambda_{1}+\lambda_{2} \mathrm{Q}^{\mathrm{FS}}+\lambda_{3} \mathrm{Q}^{\mathrm{IW}}+\lambda_{4} \mathrm{Q}^{\mathrm{AW}},
\end{equation}
where  $\lambda_{1},\lambda_{2},\lambda_{3}$ and $\lambda_{4}$ are the optimized parameters. The mapping result is also presented in Table~\ref{tab:tidres}, which achieves the best results on the tailored TID2013 database. By means of the mapping function, the pseudo-MOSs of our QUOTE-LOL database finally can be generated. In Fig.~\ref{fig:Samples}, the pseudo-MOS of each enhanced image is  annotated, from which we can observe that our database covers a wide range of quality levels. In addition, very few enhanced images achieve a pleasant visual quality. The proposed database sheds a light on how to develop better enhancement systems.

\subsection{Loop Closed Optimization}
As shown in Fig.~\ref{fig:arc},  we learn to enhance the low-light images based on a full operation chain that transfers the knowledge between quality assessment and enhancement. In particular, the NR-IQA model is learned from the annotated samples and integrated with a style modulation. Subsequently, we aim at learning to enhance the images driven by our IQA model, and finally the loop of quality optimization and assessment can be achieved with the mapping scores of three FR-IQA models as the pseudo-MOS annotations.

\subsubsection{From Enhancement to Assessment}
Benefiting from the proposed QUOTE-LOL database, a deep learning based IQA measure dedicated to enhanced low-light images is developed. It has been known that the content and the style of an image can be captured by the deep neural networks (DNNs) and are somehow separable. In particular, the convolutional feature statistics (mean, variance/std, \textit{etc.}) usually encode the style of an image. Inspired by the style transfer network, we design our NR-IQA model based on the widely used VGG16 \cite{simonyan2014very} network. The VGG16 is pretrained on the ImageNet \cite{deng2009imagenet} with its extracted features contain the  content information. 
When it comes to quality optimization, we fix all parameters of VGG16. 
The image quality is measured and regressed only by the evaluation of feature statistics (mean and std) of different layers. We find such style-sensitive architecture can highly benefit the content preservation of the final enhanced results which has been further verified in our experiments. The detailed architecture of our NR-IQA network is shown in Fig.~\ref{fig:arc}.

Given an image \textit{$I$}, it passes through the pretrained VGG16 network for multi-scale feature extraction. To encode the style information, the features in two convolution stages (the third layer $Conv3$ and the fifth layer $Conv5$) are aggregated with a global average (mean) pooling and standard deviation (std) pooling.  As shown in Fig.~\ref{fig:arc}, 
following those feature statistics, a style modulation subnetwork is introduced. In this subnetwork, for each mean/std feature, we adopt two fully-connected (FC) layers with the units set as 128 and 64 to reduce their dimensions. Subsequently, the processed mean and std features of each layer are combined by a bilinear operation. Such bilinear fusion aims to capture the pairwise interactions between each channel. Compared with concatenating the mean and std features directly, in our experiments, we find the bilinear operation leads to better quality modeling and achieves more pleasant results.
After the mean and std features are combined, three FC layers with the units set as 256, 64 and 1 are utilized for layer-wise quality prediction which is denoted as \textit{$Q_{l_1}$} and \textit{$Q_{l_2}$} in  Fig.~\ref{fig:arc}. To capture the multi-scale information, we further concatenate the processed features of the above two layers for the final quality (denoted as \textit{$Q_o$}) regression. In summary, the multi-scale quality regressions are optimized  as follows,

\begin{equation}\label{mse}
\mathcal{L}_{reg}= \frac{1}{N}
\sum_{i=1}^{N}(\left\|\textit{$Q_p - Q_o$}\right\|_{1} + \left\|\textit{$Q_p - Q_{l_1}$}\right\|_{1} + \left\|\textit{$Q_p - Q_{l_2}$}\right\|_{1}),
\end{equation}
where $\left\|\cdot\right\|$ represents the L1-norm, $i$ is the image index in a batch  with the batch size is set to $N$.

\subsubsection{From Assessment to Enhancement}
The general methodology of low-light enhancement is to recover the images with the Maximum A Posteriori (MAP) framework, based on the simultaneous maintenance of fidelity and alignment of the scene statistics. Following this vein, a fidelity and quality compatible joint loss is proposed in our optimization framework.  As we discussed in the motivation, the similarity measures of image structure and hue are incorporated in our proposed fidelity loss:
\begin{equation}\label{losscom1}
\begin{aligned}
\mathcal{L}_{f} & =  \mathbf{S} \left(\mathbf{x}^{r}, \mathbf{x}^{e}\right)+\mathbf{H} \left(\mathbf{x}^{r}, \mathbf{x}^{e})\right) \\
& = \frac{2\sigma_{\mathbf{x}^{r} \mathbf{x}^{e}}+c}{\sigma_{\mathbf{x}^{r}}^{2}+\sigma_{\mathbf{x}^{e}}^{2}+c} +  \left\|h_{\mathbf{x}^{r}} - h_{\mathbf{x}^{e}}\right\|_{1},
\end{aligned}
\end{equation}
where $\mathbf{x}^{r}$ and $\mathbf{x}^{e}$ represent the reference (normal-light) image and the enhanced image, respectively. $\sigma_{\mathbf{x}^{r} \mathbf{x}^{e}}$ is the covariance  of $\mathbf{x}^{r}$ and $\mathbf{x}^{e}$,  $\sigma_{\mathbf{x}^{r}}$ is the variance  of $\mathbf{x}^{r}$ and $\sigma_{\mathbf{x}^{e}}$ is the variance of $\mathbf{x}^{e}$.  $h_{\mathbf{x}^{r}}$ and $h_{\mathbf{x}^{e}}$ represent the hue of $\mathbf{x}^{r}$ and  $\mathbf{x}^{e}$, respectively. Regarding the quality loss, we adopt L1-norm to measure the quality difference between $\mathbf{x}^{r}$ and $\mathbf{x}^{e}$,
\begin{equation}\label{losscom2}
\mathcal{L}_{q} = \left\|Q_{max} - Q_{\mathbf{x}^{e}} \right\|_{1},
\end{equation}
where $Q^{max}$ is the maximum quality score the IQA model can achieve. In summary, our joint loss can be defined as follows,
\begin{equation}\label{losscom}
\mathcal{L} = \mathcal{L}_{f} +\lambda_5 \mathcal{L}_{q}.
\end{equation}
Herein, the $\lambda_5 $ is a hyper-parameter to adjust the importance of our quality measure. 

To reflect the underlying performance of our framework, we adopt the general architecture EDSR \cite{lim2017enhanced} as the backbone of our enhancement model. In particular, we replace the original loss functions used in EDSR  with the loss function $\mathcal{L}$ defined in Eqn.~\eqref{losscom}. 
After the enhancement model is learned, the IQA model can be renewed by model finetuning with the enhanced results and the loop will be performed iteratively until the number of iterations reaches a maximum value. 






    

\section{Experiments}

\begin{figure*}[htb]
\begin{center}
\includegraphics[width=1.0\textwidth]{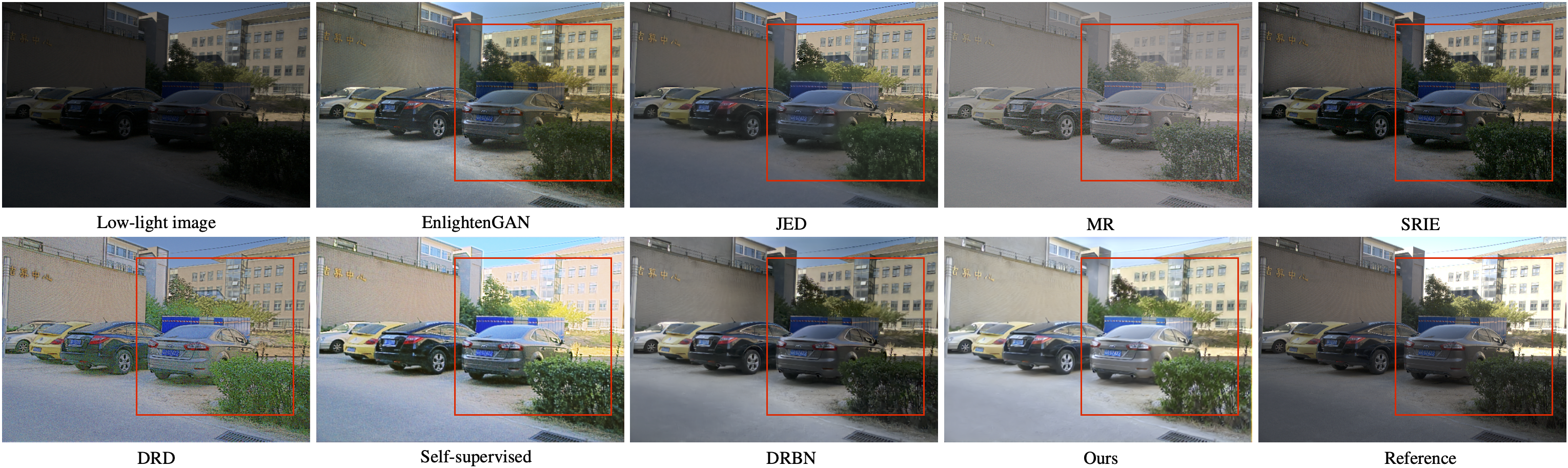}
\caption{Visual quality comparisons of images  enhanced by the proposed method and existing enhancement methods. }
\label{fig:SOTA_VS_OURS_one_content}
\vspace{-10pt}
\end{center}
\end{figure*}

\begin{figure}[htb]
\begin{center}
\includegraphics[width=1.0\textwidth]{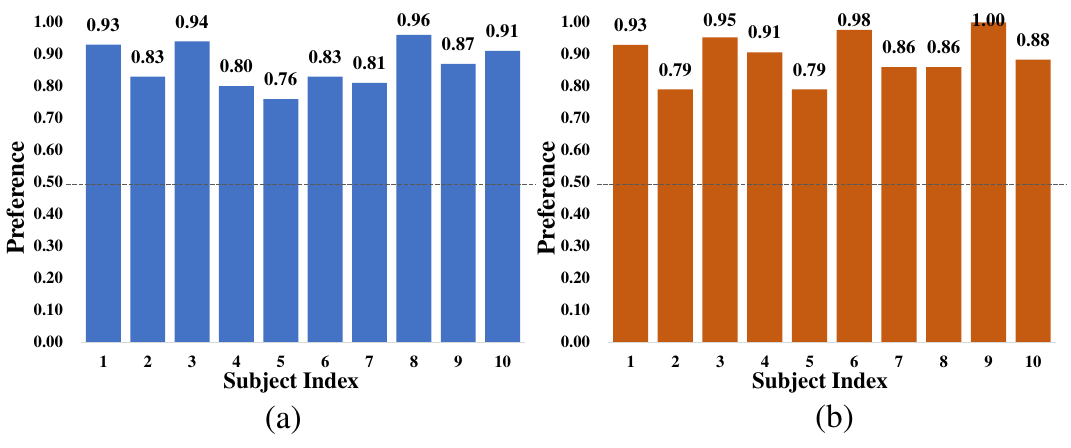}
\caption{Subjective testing results on the testing set. The percentage in favor of the proposed model is provided. (a) Comparison results between our method and the method achieving the highest pseduo-MOS in  QUOTE-LOL.  (b) Comparison results between our method and SSIM optimization.}
\label{fig:probability of preference}
\vspace{-5pt}
\end{center}
\end{figure}




\subsection{Experimental Settings}
Regarding learning of enhancement model $f_E$ , we follow the default data splitting strategy in the LOL database~\cite{wei2018deep} for generating the training and testing sets. 
Regarding the learning of NR-IQA model $f_Q$, we split the constructed QUOTE-LOL into the training set and testing test according to the content, which ensuring no content overlapping with the enhancement model.
We augment the images by randomly cropping the images to $256\times 256$. 
The  batch size  is  32  and  we  adopt Adam optimizer for optimization with the learning rate set to 1e-4. 
In total, 200 and 100 epochs are allowed during training $f_E$ and $f_Q$.  After 200 epochs, we begin to train the closed-loop framework. The learning rate drops by 0.5, when the models are fine-tuned and up to 15 and 50 epochs $f_E$ and $f_Q$ optimized at each loop. The mapping parameters $\lambda_1 \sim \lambda_4 $ in Eqn.~\eqref{map} are set to -1.8041, 2.6152, -0.2776, and 0.2563, respectively. The hyper-parameters $c$ and $\lambda_5$ in Eqn.~\eqref{losscom1} and Eqn.~\eqref{losscom} are set to 9e-4, 1.0, respectively. The max loop times we set is 10. Considering the trade-off between time consumption and performance, we adopt the third iteration outputs as our final results.

\subsection{Enhancement Results}
To verify the effectiveness of our proposed method, we compare our enhanced results with existing methods. As shown in Fig.~\ref{fig:SOTA_VS_OURS_one_content}, it is easy to find that our result tends to produce normal illumination and correct color deviation, while most of the existing methods fail to reconstruct a pleasant visual result especially in the area inside the red box. The distortions like overexposure, noise amplification and color bias are obviously introduced, while our method can adaptively restore the details according to the contexts, achieving a better balance of light enhancement and contrast modulation. From the Fig.~\ref{fig:SOTA_VS_OURS_one_content}, we can also observe that even the reference image shows an unpleasant result while our method exhibits a high potential in outperforming the reference with the quality measure introduced. We provide more comparison results in our supplementary.
For a quantitative study of our method, we have not adopted the wide-used fidelity measures like PSNR or SSIM, as they are limited by the quality of reference. Instead, we perform the subjective study on the enhanced results.  In particular, the two-alternative forced choice (2AFC) method is adopted. Given two enhanced images in each trial, subjects are forced to select the one that has better visual quality. The 2AFC method has been widely adopted in  psychophysical studies due to the discrimination capability, leading to more reliable quality labels. In our subjective study, we invited 10 inexperienced subjects for quality preference collection. In each trial, for a specific scene, our enhanced result and the result achieves the highest pseudo-MOS in QUOTE-LOL  are compared and selected by each subject. The 2AFC testing is performed on all scenes in the test set and the average individual preferences are shown in  Fig.~\ref{fig:probability of preference} (a), 
revealing a huge leap in quality preference our method achieves. 

\section{Conclusions}
In this paper, we propose the closed-loop framework for quality assessment and perceptual optimization of low-light image enhancement. 
Experimental results show the proposed framework is able to continuously improve the image quality and achieve high perceptual pleasant results. 
The generalization capability of our IQA measure is further verified in the unsupervised manner. 
As one of the first attempts on this research topic, the proposed scheme could be further improved by quality labels annotated by humans, and the interaction between human opinion and model learning opens up new innovative space for future exploration.

\begin{credits}
\subsubsection{\ackname}This work was supported by the National Natural Science Foundation of China under Grant 62401214.
\end{credits}

\bibliographystyle{splncs04}
\bibliography{new_refer}

\end{document}